\documentclass[aps,prl,showpacs,nofootinbib,superscriptaddress,twocolumn,amsmath]{revtex4-1}
\usepackage{graphicx,amssymb}
\usepackage{soul,bbm,dcolumn,booktabs}
\usepackage{multirow,array,float,enumitem}
\usepackage{lmodern,dsfont,microtype,nicefrac,microtype}
\usepackage{rotating,adjustbox,bm,xparse}
\usepackage[dvipsnames]{xcolor}
\usepackage[utf8]{inputenc}
\usepackage{upgreek}
\usepackage{array}
\usepackage{booktabs}

\newcolumntype{C}{>{$}c<{$}}
\AtBeginDocument{
\heavyrulewidth=.08em
\lightrulewidth=.05em
\cmidrulewidth=.03em
\belowrulesep=.65ex
\belowbottomsep=0pt
\aboverulesep=.4ex
\abovetopsep=0pt
\cmidrulesep=\doublerulesep
\cmidrulekern=.5em
\defaultaddspace=.5em
}
\newcommand{\be}{\begin{eqnarray}}
\newcommand{\ee}{\end{eqnarray}}

\def\fm {\mathop{\hbox{fm}}}
\def\MeV {\mathop{\hbox{MeV}}}

\def\beq{\begin{equation}}
\def\eeq{\end{equation}}
\def\beqs#1\eeqs{\beq\begin{split} #1 \end{split}\eeq}

\def\comment#1{}

\def\opbraket#1#2#3{ \left\langle #1 \left| #2 \right| #3 \right\rangle}
\def\opbraketfix#1#2#3{ \langle #1 | #2 | #3 \rangle}
\def\av#1{ \left\langle #1 \right\rangle }

\newcolumntype{L}{>{$}l<{$}} % math-mode version of "l" column type
\newcolumntype{S}{>{\footnotesize $}l<{$\normalsize}} % math-mode version of "l" column type
\def\*#1{\mathbf{#1}}

\newcolumntype{R}[2]{%
    >{\adjustbox{angle=#1,lap=\width-(#2)}\bgroup}%
    c%
    <{\egroup}%
}
\usepackage[colorlinks=true,backref=false,linktocpage=true,
pdfpagemode=UseOutlines]{hyperref}
\hypersetup{%
  bookmarksnumbered=true,
  linkcolor=NavyBlue,
  citecolor=NavyBlue,
  urlcolor=NavyBlue,
  pdftitle = {},
  pdfsubject = {},
  pdfauthor = {},
  pdfkeywords = {}
}
\begin{document}
\title{Finite-volume energy spectrum of the \texorpdfstring{$K^-K^-K^-$}{} system}

\author{Andrei~Alexandru}
\email{aalexan@gwu.edu}
\affiliation{The George Washington University, Washington, DC 20052, USA}
\affiliation{Department of Physics, University of Maryland, College Park, MD 20742, USA}
\author{Ruair\'{i}~Brett}
\email{rbrett@gwu.edu}
\affiliation{The George Washington University, Washington, DC 20052, USA}
\author{Chris~Culver}
\email{C.Culver@liverpool.ac.uk}
\affiliation{Department of Mathematical Sciences, University of Liverpool, Liverpool L69 7ZL, United Kingdom}
\author{Michael~D\"oring}
\email{doring@gwu.edu}
\affiliation{The George Washington University, Washington, DC 20052, USA}
\affiliation{Thomas Jefferson National Accelerator Facility, Newport News, VA 23606, USA}
\author{Dehua~Guo}
\email{dehuaguo@gmail.com}
\affiliation{The George Washington University, Washington, DC 20052, USA}
\author{Frank~X~Lee}
\email{fxlee@gwu.edu}
\affiliation{The George Washington University, Washington, DC 20052, USA}
\author{Maxim~Mai}
\email{maximmai@gwu.edu}
\affiliation{The George Washington University, Washington, DC 20052, USA}
%
%%%%%%%%%%%%%%%%%%%%%%%%%%%%%%%%%%%%%%%%%%%%%%%%%%%%%%%%%%%%%

\begin{abstract}
The dynamics of multi-kaon systems are of relevance for several areas of nuclear physics. However, even the simplest systems, two and three kaons, are hard to prepare and study experimentally. Here we show how to extract this information using first-principle lattice QCD results. We (1) extend the relativistic three-body quantization condition to the strangeness sector, predicting for the first time the excited level finite-volume spectrum of three kaon systems at maximal isospin, and (2)
present a first lattice QCD calculation of the excited levels of this system in a finite box. We compare our predictions with the lattice results reported here and with previous ground state calculations and find very good agreement.
\end{abstract}

\maketitle

%%%%%%%%%%%%%%%%%%%%%%%%%%%%%%%%%%%%%%%%%%%%%%%%%%%%%%%%%%%%%%%%%%%%%%%%%
\section{Introduction}\label{sec:intro}
%%%%%%%%%%%%%%%%%%%%%%%%%%%%%%%%%%%%%%%%%%%%%%%%%%%%%%%%%%%%%%%%%%%%%%%%%

In recent years hadron-hadron scattering information from first-principles lattice QCD (LQCD) has become significantly more accessible. 
Fostered by advances in theoretical and computational tools a large number of high-precision studies have been performed in the meson sector, see for example Refs.~\cite{Feng:2010es,Beane:2011sc,Aoki:2011yj,Lang:2011mn,Dudek:2012xn,Dudek:2012gj,Pelissier:2012pi,Mohler:2012na,Lang:2012sv,Mohler:2013rwa,Prelovsek:2013ela,Feng:2014gba,Wilson:2014cna,Bali:2015gji,Helmes:2015gla,Wilson:2015dqa,Orginos:2015aya,Guo:2016zos,Liu:2016cba,Bulava:2016mks, Moir:2016srx,Morningstar:2017spu,Helmes:2017smr, Alexandrou:2017mpi, Guo:2018zss,Brett:2018jqw,Andersen:2018mau,Culver:2019qtx, Mai:2019pqr,Fischer:2020fvl}, and Ref.~\cite{Briceno:2017max} for a review.
Several research areas of nuclear physics benefit from these studies. For instance, the study of pion, kaon, and proton correlations in heavy ion collisions by the ALICE@CERN collaboration~\cite{Adam:2015vja} relies on the value of the $K^-K^-$ scattering length determined in a lattice calculation~\cite{Beane:2007uh}. There are however only few results focusing on the strange sector~\cite{Beane:2007uh,Sasaki:2013vxa,Helmes:2017smr}, in contrast to the pion systems explored extensively by a number of collaborations~\cite{Culver:2019qtx,Bulava:2016mks,Dudek:2012gj,Sasaki:2013vxa,Yamazaki:2004qb,Beane:2005rj,Beane:2007xs,Feng:2009ij,Yagi:2011jn,Fu:2013ffa,Helmes:2015gla}. %%%%
%%%
Furthermore, information about many-$K^-$ systems is relevant for the understanding of strange nuclear matter and its implications to the equation of state of neutron stars. In particular, it is well known that ultra dense environments (such as those in the core of neutron stars) allow for an appearance of kaon condensates~\cite{Kaplan:1986yq,Li:1997zb,Pal:2000pb,Lee:1996ef}, that can soften the equation of state of neutron stars~\cite{Li:1997zb,Pal:2000pb,Lonardoni:2014bwa,Hell:2014xva}. Further details on the antikaon interaction with baryonic matter can be found in reviews~\cite{Gal:2016boi,Mai:2020ltx}.

Today, the frontier of hadronic scattering in LQCD is in the scattering of three mesons. Pioneering lattice calculations have moved from the extraction of the ground states of such systems~\cite{Beane:2007es,Detmold:2008fn,Detmold:2008yn} to the high-precision determination of multiple excited three-hadron states~\cite{Horz:2019rrn,Woss:2019hse,Culver:2019vvu,Fischer:2020jzp,Hansen:2020otl}. 
Significant progress has also been made in the development of formalisms relating the finite- and infinite-volume three-hadron spectrum~\cite{Hansen:2020zhy,Mai:2019fba,Pang:2020pkl,Blanton:2020jnm,Blanton:2020gha, Blanton:2019vdk,Mai:2018djl, Guo:2019hih,Romero-Lopez:2019qrt,Mai:2018djl,Zhu:2019dho,Guo:2018ibd, Doring:2018xxx,Guo:2018xbv,Romero-Lopez:2018rcb,Klos:2018sen, Mai:2017bge,Guo:2017crd, Guo:2017ism, Hammer:2017kms,Briceno:2018aml,Briceno:2018mlh,Briceno:2017tce,Guo:2016fgl,Hansen:2016fzj,Hansen:2015zga,Jansen:2015lha,Hansen:2014eka,Polejaeva:2012ut,Roca:2012rx,Briceno:2012rv,Bour:2012hn, Kreuzer:2012sr,Kreuzer:2009jp,Kreuzer:2008bi, Meng:2017jgx, Hammer:2017uqm, Meissner:2014dea, Bour:2011ef, Kreuzer:2010ti,Guo:2020kph,Guo:2020ikh}.
Applications of such approaches to LQCD data have thus far been for three pion systems in maximal isospin~\cite{Mai:2018djl,Mai:2019fba,Culver:2019vvu,Blanton:2019vdk,Fischer:2020jzp,Hansen:2020otl}. %3pi applications

In this paper, we extend these methods to explore a new area:
we present both the first determination of the excited three-kaon finite-volume spectrum from LQCD, along with the first connection to infinite-volume scattering using the formalism of Refs.~\cite{Mai:2019fba, Mai:2019pqr, Mai:2018djl, Doring:2018xxx, Mai:2017bge}. The latter is extended to the three-flavor sector allowing for chiral extrapolations along arbitrary~$M_K(M_\pi)$ trajectories using constraints from chiral symmetry. Such implementations are standard in the two-body sector~\cite{Niehus:2020gmf,Rendon:2020rtw,Molina:2020qpw,Hu:2017wli,Nebreda:2010wv,Pelaez:2010fj,Nebreda:2011di,Guo:2016zep}, but not yet explored for the three-body systems. 
The present study closes this gap, using relativistic three-body formalism implementing two-body input from the inverse amplitude approach~\cite{GomezNicola:2001as,Truong:1988zp}.

%%%%%%%%%%%%%%%%%%%%%%%%%%%%%%%%%%%%%%%%%%%%%%%%%%%%%%%%%%%%%%%%%%
\section{Finite-volume spectrum from Lattice QCD}\label{sec:lattice}
%%%%%%%%%%%%%%%%%%%%%%%%%%%%%%%%%%%%%%%%%%%%%%%%%%%%%%%%%%%%%%%%%%

The finite-volume spectrum of hadronic states can be directly accessed by studying correlation functions in the framework of LQCD.  
Here we review the procedure for extracting the finite-volume spectrum of $K^-K^-K^-$.
The energy levels of hadrons in a finite volume can be extracted from the large time behavior of correlation functions consisting 
of interpolating operators, $\mathcal{O}_i$, which create/annihilate the hadrons of interest,
\beq
    C_{ij}(t)=\av{\mathcal{O}_i(t)\mathcal{O}_j^{\dagger}(0)}=\sum_n \opbraket{0}{\mathcal{O}_i}{n}\opbraketfix{n}{\mathcal{O}_j^{\dagger}}{0}e^{-E_n t}.
\eeq
If the operators are constructed to overlap with the states $n$ of interest, we can extract the finite-volume energies $E_n$.  An important tool to allow the extraction of multiple finite-volume energies is to perform a variational analysis on a matrix of correlation functions constructed from several operators.  This is equivalent to solving a generalized eigenvalue problem~\cite{Luscher:1990ck,Michael:1982gb,Blossier:2009kd}, and extracting the finite-volume spectrum from the eigenvalues of the correlation matrix.  Due to the precision with which we can measure the correlation functions, thermal effects due to the finite temporal extent must be accounted for as in Ref.~\cite{Culver:2019vvu}.

%%%%%%%%%%%%%%%%%%%%%%%
%%%%%%%%%%%%%%%%%%%%%%%
\begin{table}[t]   
\resizebox{\columnwidth}{!}{
\begin{tabular}{llll l}
\toprule
Label~~&$N_t\times N^3 $&$a[\fm]$&$N_\text{cfg}$&\\
\midrule
$\mathcal{E}_1$&$48\times24^3$&$0.1210(2)(24)$&$300$
&$aM_{\pi}=0.1931(4)$\\
&&&
&$af_{\pi}~=0.0648(8)$\\
&&&
&$aM_{K}=0.3236(3)$\\
&&&
&$af_{K}~=0.1015(2)$\\
\midrule
$\mathcal{E}_4$&$64\times24^3$&$0.1215(3)(24)$&$400$
&$aM_{\pi}=0.1378(6)$\\
&&&
&$af_{\pi}~=0.0600(10)$\\
&&&
&$aM_{K}=0.3132(3)$\\
&&&
&$af_{K}\,=0.0980(2)$\\
\bottomrule
\end{tabular} } 
\caption{
\label{table:gwu_lattice}
Details and results of the GWUQCD $N_f=2$ ensembles used in this study.  Here $a$ is the lattice spacing,  $N_{\text{cfg}}$ the number of Monte Carlo configurations for each ensemble, and  $a M_{\pi}$ and $a M_K$  the pion and kaon masses, respectively. The errors in the parenthesis are stochastic. For the lattice spacing we also include an estimate for the systematic error of~2\%.}  
\end{table}
%%%%%%%%%%%%%%%%%%%%%%%
%%%%%%%%%%%%%%%%%%%%%%%

The overlap factor $\opbraket{n}{\mathcal{O}_j}{0}$ is non-zero only if our operators and states $n$ have the same quantum numbers.  In a finite cubic volume, the rotational symmetry group is reduced from $SO(3)$ to $O_h$.  We have to therefore construct our operators with definite quantum numbers according to the irreducible representations (irreps) of $O_h$.  An important consequence is that the irreps of $O_h$ mix different angular momentum from the infinite volume.  The symmetry is further reduced if the system is studied with non-zero total momentum.

To create operators which overlap with the three-kaon spectrum, we begin by constructing a single kaon interpolator according to 
\beq
    K^-(\Gamma(\bm{p}),t)=\bar{u}(t)\Gamma(\bm{p})s(t),
\eeq
where $s, u$ are the quark fields, and the momentum matrix $\Gamma(\bm{p})=e^{i\bm{p}\cdot\bm{x}}\gamma_5$ projects the operator to  definite momentum.  Our three-kaon operators are now just a product of three single kaon operators.  We project the three kaon operators to irreps of the cubic group.  To project to row $\lambda$ of irrep $\Lambda$ of group $G$, we evaluate
\beqs
 {\cal O}_{K_1K_2K_3}=\sum_{g\in G}&U^{\Lambda}_{\lambda\lambda}(g)\text{det}(R(g)) \\
 \times\,&K^-(R(g)\bm{p}_1)K^-(R(g)\bm{p}_2)K^-(R(g)\bm{p}_3),
 \label{equation:project_op}
\eeqs
where $p_1,p_2,p_3$ are the three-momenta of each kaon, $R$ is the three-dimensional rotation matrix associated with $g$, and $U$ is the representation matrix of $g$ in irrep $\Lambda$.

The GWUQCD ensembles are generated using two mass-degenerate light quarks ($N_f=2$ QCD), using the nHYP-smeared clover action. Lattice parameters of the ensembles used here are listed in Table~\ref{table:gwu_lattice}. Details of the ensemble generation, including some discussion on tuning the bare strange quark mass, can be found in Refs.~\cite{Guo:2016zos,Niyazi:2020erg}.
The pion and kaon decay constants, $f_\pi$, $f_K$ are determined using the procedure outlined in Ref.~\cite{Fritzsch:2012wq}.
The strange quark mass is tuned by setting the ratio $R=(M_K/M_{\pi})^2$ to its physical value.
For the valence quarks appearing in (kaon) interpolating operators, both light and strange (all-to-all) quark propagation is treated using the LapH method~\cite{Peardon:2009gh}. The all-to-all LapH propagators were computed using our optimized inverters~\cite{Alexandru:2011ee}. The lattice results and predictions are tabulated in the Supplementary Material. Jackknife samples are provided as ancillary files with the arXiv submission.

%%%%%%%%%%%%%%%%%%%%%%%%%%%%%%%%%%%%%%%%%%%
%%%%%%%%%%%%%%%%%%%%%%%%%%%%%%%%%%%%%%%%%%%
\begin{figure}
    \centering
    \includegraphics[width=1.\linewidth]{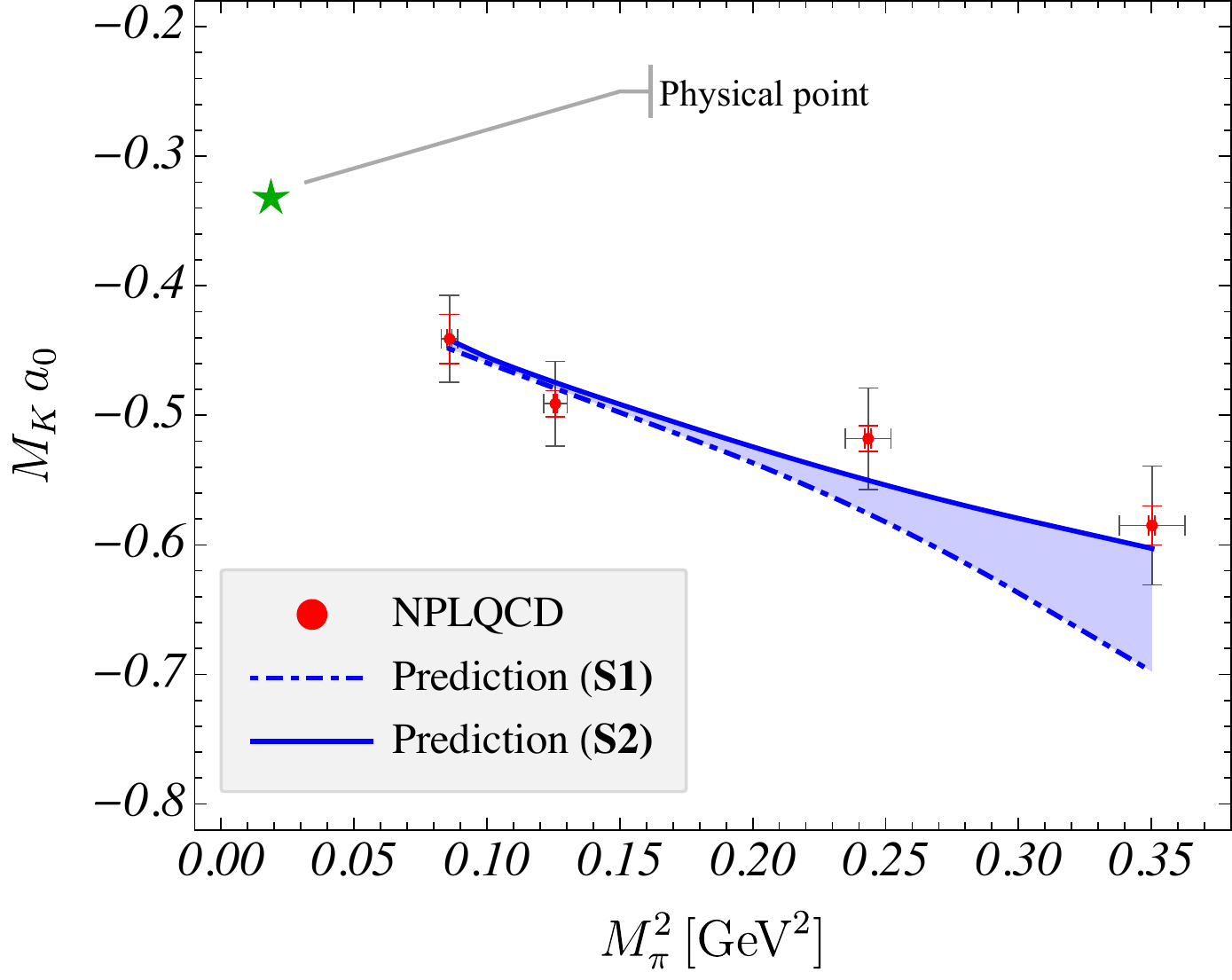}
   \caption{
    \label{fig:scatlength}
 The $I=1$ $K K$ scattering lengths from NPLQCD~\cite{Beane:2007uh} with statistical (red) and systematic (gray) error bars. The chiral prediction at the physical point is indicated (green star), as well as predictions for different pion masses using $f_\pi,\,f_K$ from NPLQCD~\cite{Beane:2006kx} (solid blue line), or NLO extrapolating $f_\pi$ (blue dash-dotted line).
 }
\end{figure}
%%%%%%%%%%%%%%%%%%%%%%%%%%%%%%%%%%%%%%%%%%%
%%%%%%%%%%%%%%%%%%%%%%%%%%%%%%%%%%%%%%%%%%%

%%%%%%%%%%%%%%%%%%%%%%%%%%%%%%%%%%%%%%%%%%%
%%%%%%%%%%%%%%%%%%%%%%%%%%%%%%%%%%%%%%%%%%%
\begin{figure*}[t]
    \centering
    \includegraphics[height=5.2cm]{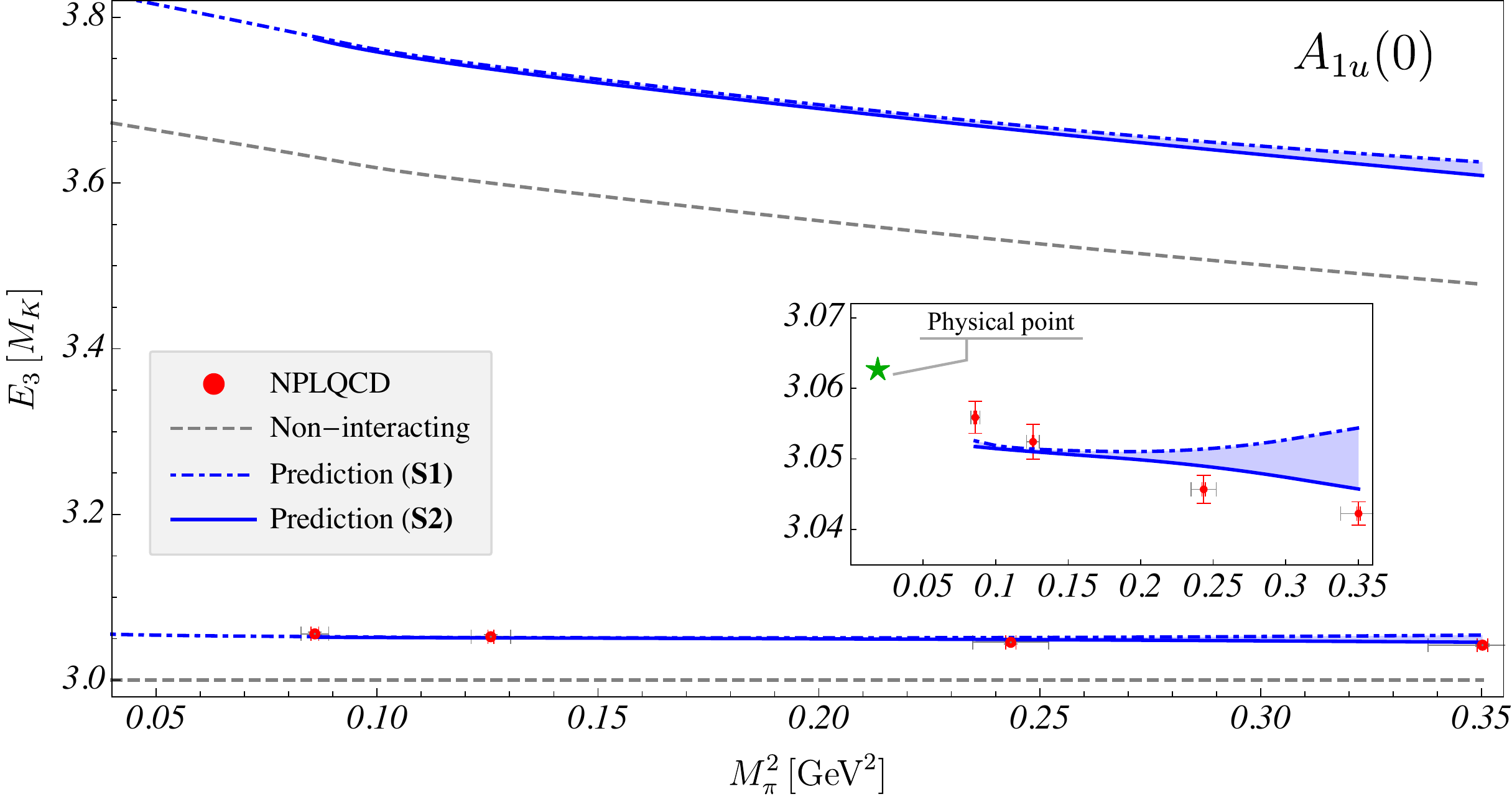}
    ~~~~~
    \includegraphics[height=5.2cm]{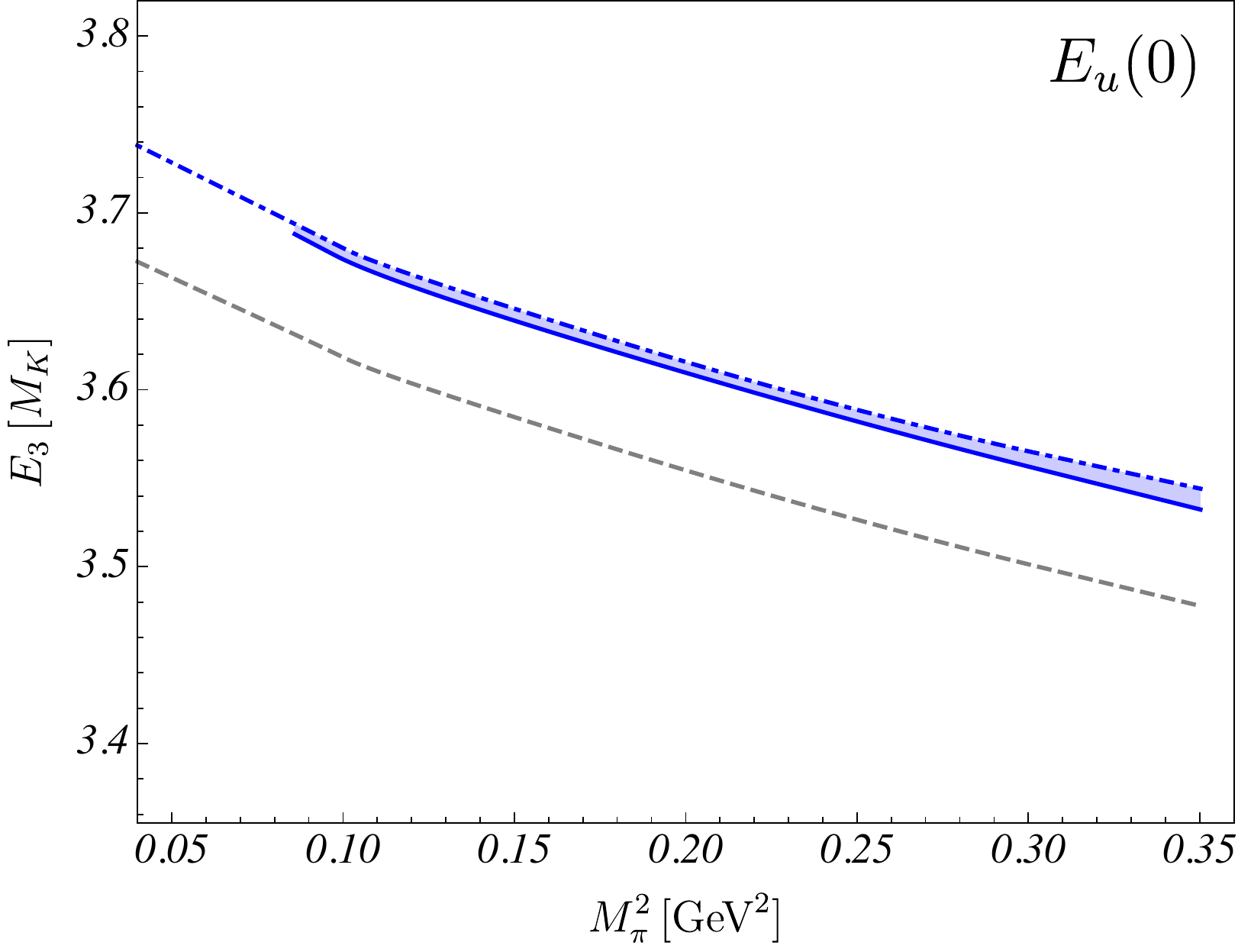}
    \\
    \includegraphics[height=4.4cm]{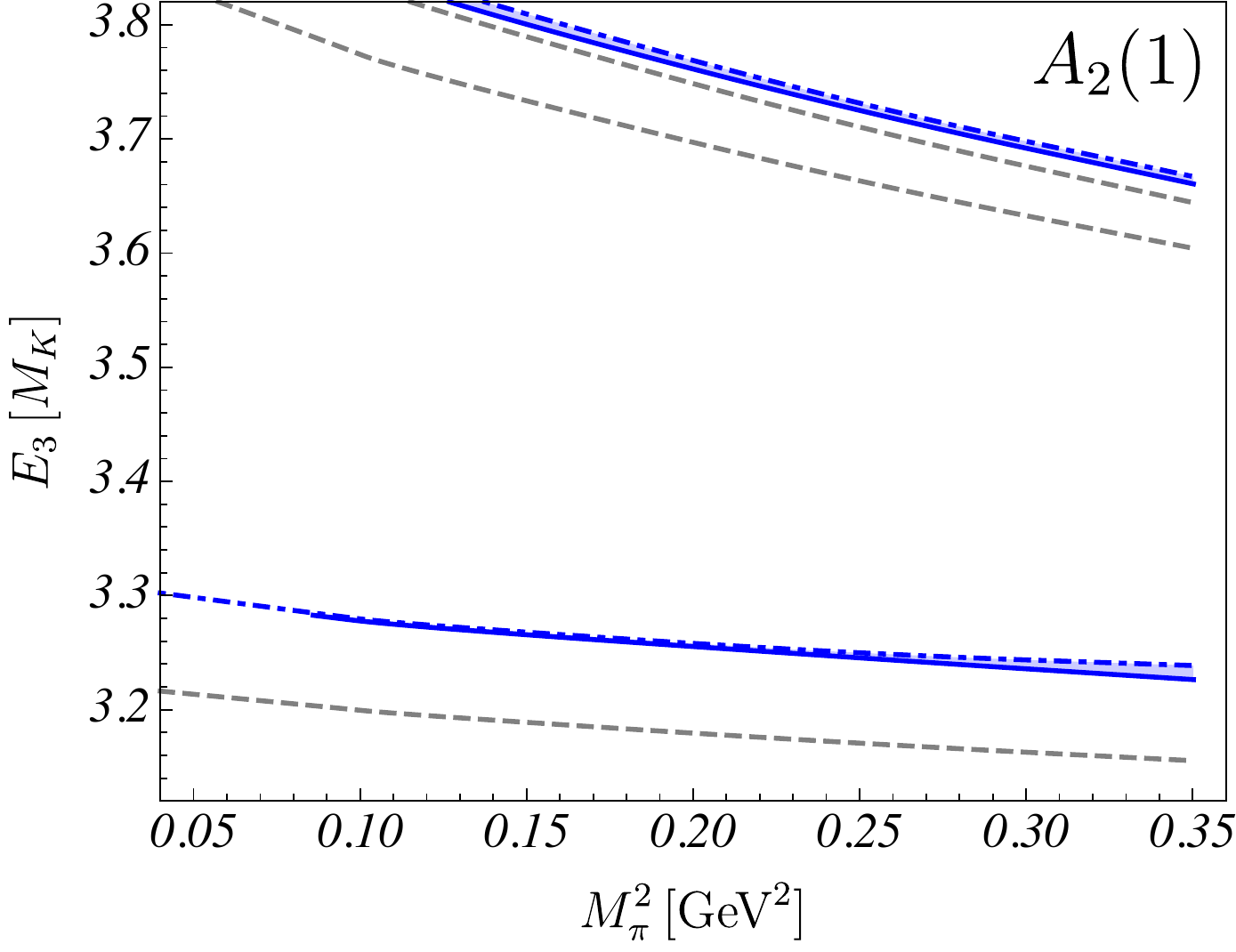}~~~
    \includegraphics[height=4.4cm,trim=1cm 0 0 0,clip]{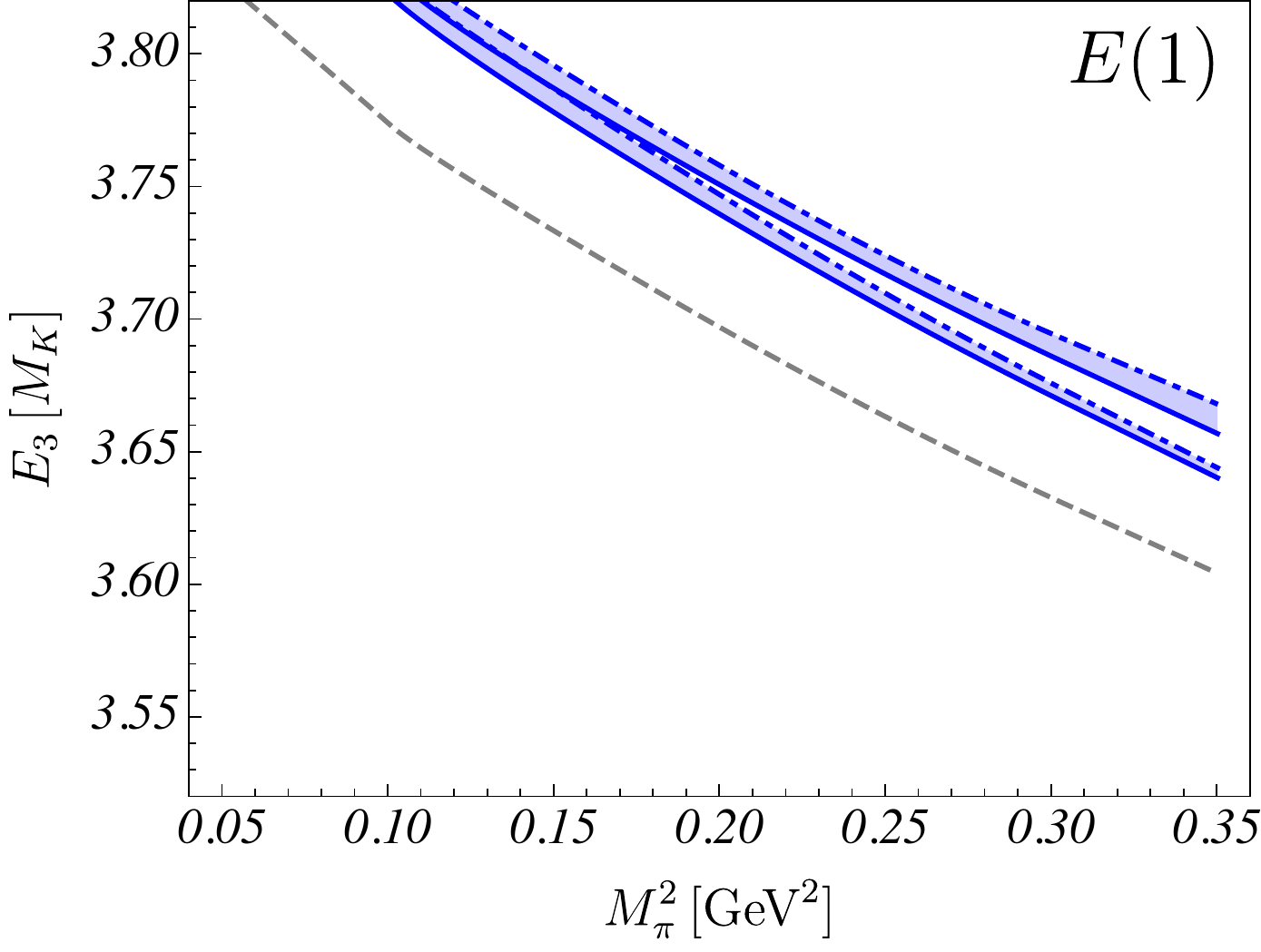}~~~
    \includegraphics[height=4.4cm,trim=1cm 0 0 0,clip]{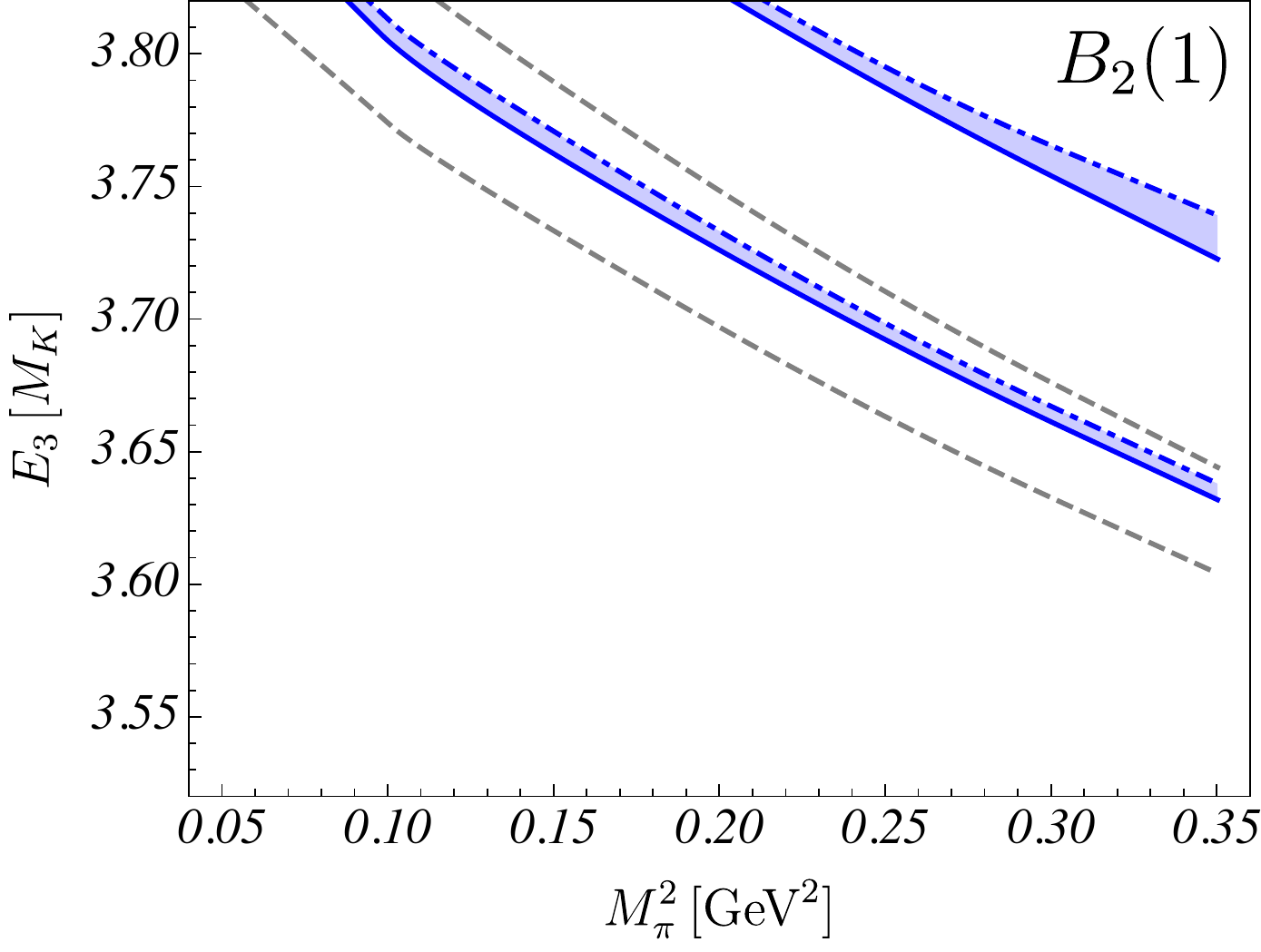}
    \caption{
    \label{fig:A1(0)NPLQCD}
    Comparison of the predicted $K^-K^-K^-$ finite-volume spectrum to the results of LQCD calculations~\cite{Detmold:2008yn,Beane:2006kx} by the NPLQCD collaboration. Top and bottom row show projections to relevant irreps for ${\bm P}={\bm 0}$ and ${\bm P}={(1,0,0)}$ cases, respectively.
    The $M_K(M_\pi)$ trajectory is chosen as in the latter references, while the decay constants are determined from the NLO chiral extrapolations (dot-dashed line) or by setting them directly to the NPLQCD values (blue solid lines). In the top left figure, the red (gray) error bars represent the uncertainty quoted in Ref.~\cite{Detmold:2008yn} (including variation of scale setting).
    The insert in the $A_{1u}$ plot shows the ground state data, predictions, and a prediction for the physical point.
    }
\end{figure*}
%%%%%%%%%%%%%%%%%%%%%%%%%%%%%%%%%%%%%%%%%%%
%%%%%%%%%%%%%%%%%%%%%%%%%%%%%%%%%%%%%%%%%%%

%%%%%%%%%%%%%%%%%%%%%%%%%%%%%%%%%%%%%%%%%%%%%%%%%%%%%%%%%%%%%%%%%%
\section{Finite-volume spectrum from infinite-volume physics}
\label{sec:truf}
%%%%%%%%%%%%%%%%%%%%%%%%%%%%%%%%%%%%%%%%%%%%%%%%%%%%%%%%%%%%%%%%%%

In the present work we utilize the three-body relativistic quantization condition (3bQC) derived in Ref.~\cite{Mai:2017bge} extended later to higher irreps~\cite{Doring:2018xxx}, elongations~\cite{Culver:2019vvu}, and boosts~\cite{Mai:2019fba}. For the $I=3/2$, $S=-3$ three-meson channel the 3bQC reads
%%%%%%%%%%%%%%%%%%%%%%%%%%%%
\begin{equation}
\begin{aligned}[t]
\det\Big[
B(E_3)&+C(E_3)\\
&+E_L
\left(
K_2^{-1}(E_3)+
\rho_L(E_3,\bm{P})
\right)
\Big]^{\Gamma}_{\bm{p}\bm{q}}=0\,,
\end{aligned}
\label{eq:3bqc}
\end{equation}
%%%%%%%%%%%%%%%%%%%%%%%%%%%%
where $E_3$ and $\bm P$ denote the center of mass energy and total three-momentum of the three-body system, respectively.
Note that the implicit dependence on the latter is suppressed. The determinant is taken with respect to the in/outgoing discrete lattice spectator momenta $\bm{p}/\bm{q}$ after projecting the elements in parenthesis to an irrep $\Gamma$. The non-diagonal matrix $B$ denotes the one-particle exchange term, while the diagonal matrix $\rho_L(E_3,\bm{P})$ represents the two-body self-energy term, see, e.g., appendix of Ref.~\cite{Mai:2019fba} for explicit expressions.  The propagation of the spectator yields the factor $[E_L]_{\bm{p}\bm{q}}=\delta_{\bm{p}\bm{q}}2L^3\sqrt{M_\pi^2+\bm{p}^2}$.

The only unknown pieces of the quantization condition are matrices  $K_2^{-1}(E_3)$ and $C(E_3)$, encoding dynamics of two- (via the usual $K$-matrix) and three-body interactions, respectively. Since not many data is available yet for the $3K^-$-system and in analogy to the similar $3\pi^+$ system~\cite{Mai:2018djl}, we set the latter to zero. 
The two-body $K$-matrix is restricted to the dominant $S$-wave, noting that due to the nature of the 3bQC all relative partial waves between the spectator and the two-body subsystem are included automatically by the one-particle exchange term $B$. Specifically, the $K$-matrix is chosen to match the inverse amplitude method~\cite{Truong:1988zp,Guerrero:1998ei,Pelaez:2004xp,Nebreda:2010wv,GomezNicola:2001as} -- a very successful description of  two-meson scattering across wide energy and meson mass ranges and all two-pseudoscalar meson interaction channels~\cite{Mai:2019pqr},
%%%%%%%%%%%%%%%%%%%%%%%%%%%%
\begin{equation}
\begin{aligned}[t]
T_2(s)=\frac{(T_{\rm LO}(s))^2}{T_{\rm LO}(s)-T_{\rm NLO}(s)}
=\frac{1}{K_2^{-1}(s)-\rho(s)}\,.
\label{eq:two-body}
\end{aligned}
\end{equation}
%%%%%%%%%%%%%%%%%%%%%%%%%%%%
Here, $T_{\rm (N)LO}$ refers to the (next-to-)leading chiral order scattering amplitudes~\cite{Gasser:1984gg}, and $\rho(s)$ denotes the usual finite part of the two-body self-energy evaluated in dimensional regularization. The $K^-K^-$ amplitude to one loop is obtained by using crossing symmetry for results of Ref.~\cite{GomezNicola:2001as}. A summary of the relevant formulas is included in the Supplementary Material. In particular, the corresponding $K$-matrix depends on $\{M_\pi,M_K,f_\pi,f_K\}$ as well as renormalized low-energy constants (LECs) $\{L_i^r\}$. The effect of the first set of parameters is more important than the latter for not too large meson masses, because the chiral series is ordered in powers of $M^2/f^2$. Thus, we fix the LECs to the results of the most recent global fits to the lattice results~\cite{Molina:2020qpw}
(discussion of older LECs is moved to the Supplementary Material),
but explore various scenarios for the remaining inputs below.

%%%%%%%%%%%%%%%%%%%%%%%%%%%%%%%%%%%%%%%%%%%
%%%%%%%%%%%%%%%%%%%%%%%%%%%%%%%%%%%%%%%%%%%
\begin{figure*}[t]
    \centering
    \includegraphics[height=5.5cm, trim=0 0 0 0,clip]{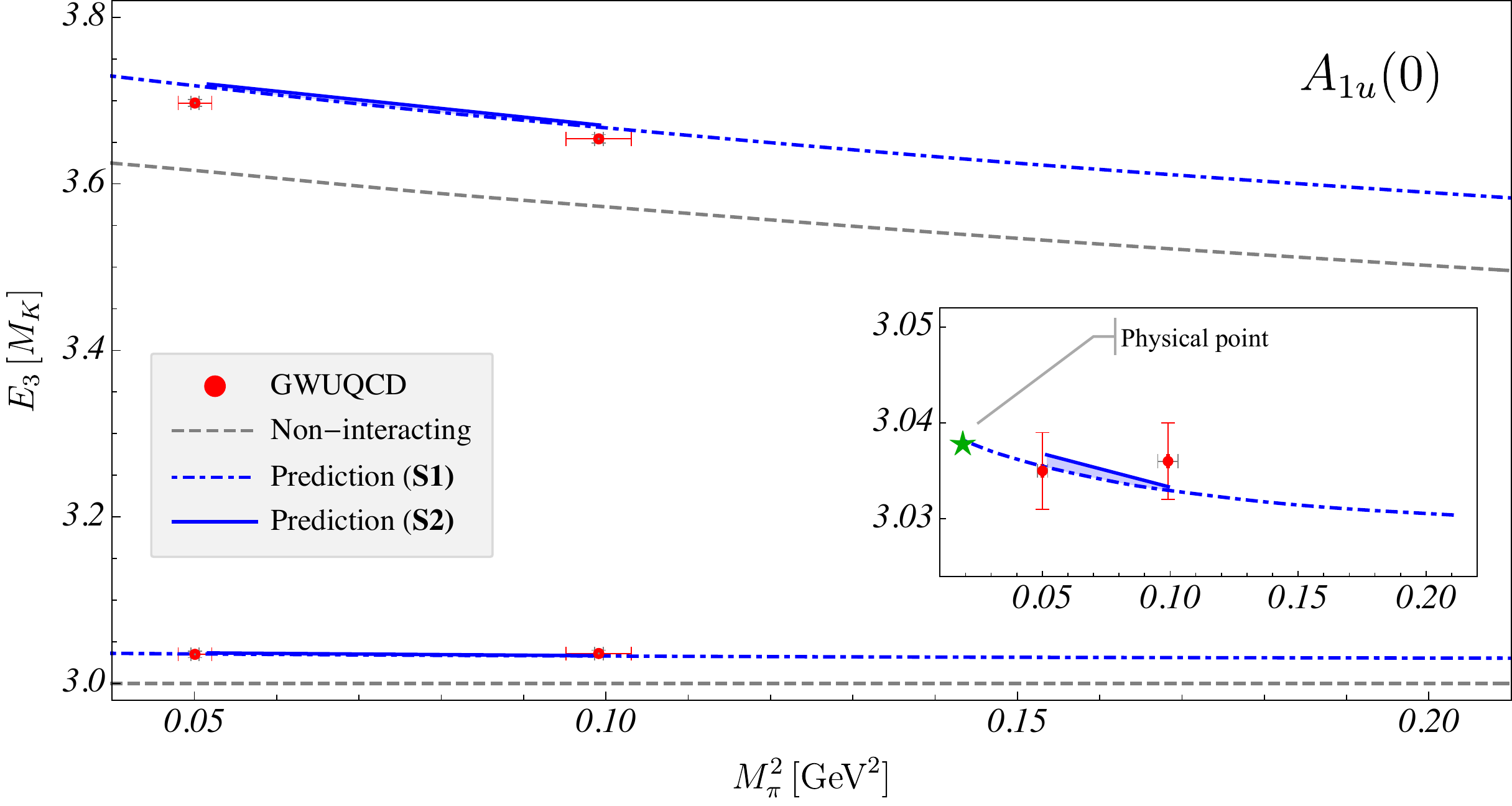}
    ~~~
    \includegraphics[height=5.5cm]{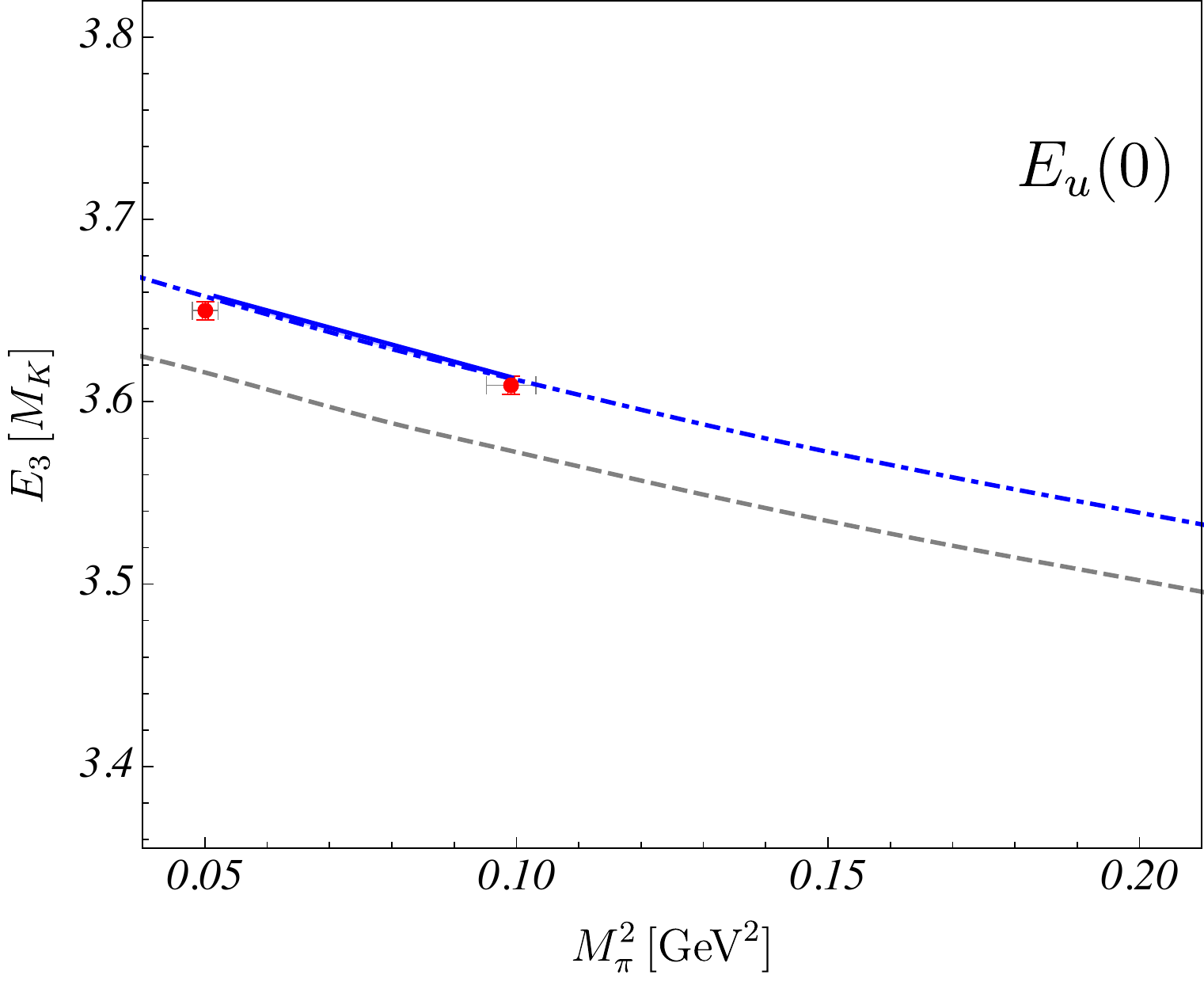}
    \caption{
    \label{fig:3BenGW}
    Comparison of the predicted $K^-K^-K^-$ finite volume spectrum to the present lattice results (red data) for ${\bm P}={\bm 0}$. 
    The $M_K(M_\pi)$ trajectory extends linearly from the physical point through the two shown lattice points.
    The insert in the $A_{1u}$ plot shows the ground state data, predictions, and a prediction for the physical point.
    Note that the excited states in $A_{1u}$ and $E_u$ are close to or beyond the $\pi KKK$ threshold, but below the lowest relevant lattice threshold, for which two kaons necessarily have finite back-to-back momenta due to parity conservation.
    }
\end{figure*}
%%%%%%%%%%%%%%%%%%%%%%%%%%%%%%%%%%%%%%%%%%%
%%%%%%%%%%%%%%%%%%%%%%%%%%%%%%%%%%%%%%%%%%%

As a check we evaluate the scattering length $M_Ka_0=T_2(4M_K^2)$ at different meson masses and compare with the NPLQCD collaboration results~\cite{Beane:2007uh} along their $(M_\pi,M_K)$ trajectory  $(M_\pi,M_K)\in\{(293,583),(355,601),(493,643),(592,680)\}$~MeV. For the decay constants we compare two scenarios: ({\bf S1}) by extrapolating the pion decay constant using input at the physical point and NLO chiral expressions~\cite{Gasser:1984gg} with LECs from Ref.~\cite{Molina:2020qpw} and ({\bf S2}) by using the meson decay constants determined on the lattice~\cite{Beane:2006kx}. These two scenarios differ only by higher chiral orders and are employed as representatives of the systematic uncertainty of our predictions. The results are depicted in Fig.~\ref{fig:scatlength}. They show that the three-flavor formulation of the inverse amplitude approach~\eqref{eq:two-body} is a perfectly suitable parametrization of the two-body dynamics at unphysical meson masses. Higher order terms yield sizable corrections above $M_\pi\approx 4M_\pi^{\rm phys}$ as expected.

Before coming to the results on three-body spectra we point out the major difference between the 3bQC and its two-body equivalent. The 3bQC remains a determinant equation even for the simplest one-channel case. For a fixed energy and momentum of the three-body system, the two-body input is required for a large kinematic range ($s$ in Eq.~\eqref{eq:two-body}) due to the variable spectator momentum. Therefore, the two-body amplitude is often evaluated for subthreshold values of $s$.
Various approaches to this issue have been studied in the past~\cite{Mai:2017bge,Mai:2019fba} and it was found that the obtained finite-volume spectra depend little on the subthreshold region. For the present case, we confirm this observation explicitly by varying the cut in the spectator momentum space in Eq.~\eqref{eq:3bqc}. As we change this from the value used throughout this study, $L|\bm{ p}_{\text{max}}|=2\pi \sqrt{5}$, to $L|{\bm{p}}_{\text{max}}|=2\pi \sqrt{11}$, the largest change $(\sim3\times 10^{-3}~\%)$ among all levels in the GWUQCD setup for $M_\pi=315$~MeV happens for the first excited level in $A_{1u}$.
Similarly, we study the dependence on the subthreshold $KK$ amplitude by replacing $K_2^{-1}(s)$ with a real-valued constant at $s=3M_K^2$ and then at $s=3.95M_K^2$ leading to a maximal change of any energy eigenvalue of~0.02\% which is a fraction of the smallest statistical uncertainty in the GWUQCD lattice data. The dependence of the results on the use of modified IAM (mIAM)~\cite{GomezNicola:2007qj} instead of IAM leads to $\lesssim 0.1\%$ change of the scattering length and is, therefore, of similar size for the three-body ground state energy shift. In summary, these sources of systematic uncertainty are very small.

%%%%%%%%%%%%%%%%%%%%%%%%%%%%%%%%%%%%%%%%%%%%%%%%%%%%%%%%%%%%%
\section{Comparison and discussion}\label{sec:results}
%%%%%%%%%%%%%%%%%%%%%%%%%%%%%%%%%%%%%%%%%%%%%%%%%%%%%%%%%%%%%

First, we turn to previous LQCD results, namely the ground state $A_{1u}(0)$ levels determined by the NPLQCD collaboration~\cite{Detmold:2008yn,Beane:2006kx} in a cubic box of $L=2.5$~fm at four pairs of pseudo-scalar meson masses. These results are depicted in the top left panel of Fig.~\ref{fig:A1(0)NPLQCD} and overlayed by the predictions of the 3bQC. As before, differences between scenarios {\bf S1} and {\bf S2} are visualized by the light blue band. We observe encouraging agreement between our predictions and the NPLQCD results. We also find a similar increase in the size of NNLO effects at higher pion masses, as observed in the two-body results.
Going beyond the ground state level, we extend our predictions to excited states, other irreps, and boosts (lower panel of Fig.~\ref{fig:A1(0)NPLQCD}).

New LQCD results are obtained in this work, including for the first time, excited $K^-K^-K^-$ energies, in multiple irreps. This provides an opportunity for a series of unique tests of the predicted finite-volume spectra. Following both scenarios {\bf S1} and {\bf S2} along the  chiral trajectory (see Table~\ref{table:gwu_lattice}) the predictions for the GWUQCD setup are shown in Fig.~\ref{fig:3BenGW}. The ground state is in excellent agreement with the predictions as was the case for the heavier pion mass results from NPLQCD. For the excited $A_{1u}(0)$ levels the slight tension with the prediction could be some hint of the need for a non-zero three-body force. Of course, other possible sources for the discrepancy could be (i) the chiral prediction itself is not perfect, or (ii) that the partial quenching of the strange quark plays a role. This will be investigated in a future study.

The $E_u(0)$ levels agree with the predictions well. Note that this irrep is dominated by $D$-wave. Since the two-body interaction is typically smaller for higher partial waves, the major contribution seems to come from the one-particle exchange term $B$, with no obvious need for contact terms beyond that. In fact, this is very similar to the observed $E_u^-(0)/A_{1u}^-(0)$ pattern for the three-pion system noted in Ref.~\cite{Mai:2019fba}. In both cases, the pattern confirms the dominance of the exchange contribution, which is a direct consequence of the S-matrix principle of three-body unitarity.

In summary, we have traced a pathway for studying multi-kaon systems using lattice QCD. We presented the first LQCD calculation of excited three kaon states, in multiple irreps, and at multiple pion masses. We have also extended the relativistic three-body quantization condition to the strange sector, allowing for chiral extrapolations along arbitrary trajectories. We find that this extension consistently describes the data from two independent lattice calculations of multi-kaon systems. In the long run, this provides an avenue for extracting information relevant for strange resonances, kaon condensates, and heavy-ion collisions.

%%%%%%%%%%%%%%%%%%%%%%%%%%%%%%%%%%%%%%%%%%%%%%%%%%%%%%%%%
\bigskip
\acknowledgements
This material is based upon work supported by the National Science Foundation under Grant No. PHY-2012289 and by the U.S. Department of Energy under Award Number DE-SC0016582 (MD and MM) and DE-FG02-95ER40907 (AA,FXL,RB,CC). 
RB is also supported in part by the U.S. Department of Energy and ASCR, via a Jefferson Lab subcontract No. JSA-20-C0031.  
CC is supported by UK Research and Innovation grant {MR/S015418/1}.
%%%%%%%%%%%%%%%%%%%%%%%%%%%%%%%%%%%%%%%%%%%%%%%%%%%%%%%%%

%%%%%%%%%%%%%%%%%%%%%%%%%%%%%%%%%%%%%%%%%%%%%%%%%%%%%%%%%
%%%%%%%%%%%%%%%%%%%%%%%%%%%%%%%%%%%%%%%%%%%%%%%%%%%%%%%%%
\bibliography{main.bib}
%%%%%%%%%%%%%%%%%%%%%%%%%%%%%%%%%%%%%%%%%%%%%%%%%%%%%%%%%
%%%%%%%%%%%%%%%%%%%%%%%%%%%%%%%%%%%%%%%%%%%%%%%%%%%%%%%%%

\clearpage
\appendix
\begin{onecolumngrid}

%%%%%%%%%%%%%%%%%%%%%%%%%%%%%%%%%%%%%%%%%%%%%%%%%%%%%%%%%%%%%
\section{Next-to-leading order chiral \texorpdfstring{$K^-K^-$}{} amplitude}
\label{app:NLO-chpt}
%%%%%%%%%%%%%%%%%%%%%%%%%%%%%%%%%%%%%%%%%%%%%%%%%%%%%%%%%%%%%

In the following we provide the explicit formulas for the scattering matrices of the two-body input. They rely on the three-flavor chiral Lagrangian of the leading and next-to-leading chiral order~\cite{Gasser:1984gg}. We chose the formulation with all decay constants replaced by the ``physical'' pion one, which is done consistently at the forth chiral order, i.e., differences are of the order $\mathcal{O}(p^6)$. In practice, we obtain the $K^-K^-\to K^-K^-$ amplitude from the $K^+K^-\to K^+K^-$ amplitude quoted in Ref.~\cite{GomezNicola:2001as} (see also Ref.~\cite{Guerrero:1998ei} for the original calculation) by crossing symmetry which amounts to exchanging $s\leftrightarrow u$ in the latter,
%%%%%%%%%%%%%%%%%%%%%%%%%%%%
\begin{equation}
\begin{aligned}[t]
\label{eq:amp4kch}
T(s,t,u)=&T_{\rm LO}(s,t,u)+T_{\rm NLO}(s,t,u)+\dots\\
%%Tree
=&\left[\frac{2M_K^2-s}{f_\pi^2}\right]_{\rm LO}+\\
%%muK
&\left[-\frac{ \mu_K  }{6 f_\pi^2  M_K^2} \left( 5 \left( u^2 + u t + t^2
\right)  + 6 s^2 - 13 s M_K^2 - 8  M_K^4 \right)\right.
\\
%%mspi
&~+\frac{\mu_\pi}{2 f_\pi^2} \left( 5\left( s - 2  M_K^2 \right)-
\frac{ 11 u^2 + 8 u t + 11 t^2 + 8 s  M_K^2 - 32  M_K^4  } {24
M_\pi^2} + \frac{ 9 \left( u^2 + t^2 \right)  + 24 s  M_K^2 -
 64  M_K^4  }{16( M_K^2 -  M_\pi^2)}
\right)  \\
%%mue
&~+\frac{\mu_\eta} {12 f_\pi^2}
 \left( 64  M_K^2 - 2  M_\pi^2 - 27 s
- \frac{81 \left( u^2 + t^2 \right) - 36 \left( u + t \right)
M_\pi^2 + 8  M_\pi^4}{12 M_\eta^2} +
      \frac{ 9 \left( u^2 + t^2 \right)  + 24 s   M_K^2 -
           64  M_K^4 }{ 2(M_\pi^2 -  M_\eta^2)} \right)\\
%% L1,L2,L3
&~+\frac{4}{f_\pi^4}
\left(2L_2^r(s-2M_K^2)^2+(2L_1^r+L_2^r+L_3)
\left((u-2M_K^2)^2+(t-2M_K^2)^2\right)
%% L4 a L8
-4L_4^r s M_K^2-2 L_5^r(s-2M_K^2)M_\pi^2\right.\\
&~-\left.4\left(L_5^r-2(2L_6^r+L_8^r)\right))M_K^4\right)
%% Polinomio suelto
+\frac{186 u t -177 s^2+1032 s M_K^2 -1648 M_K^4}{2304 f_\pi^4
{\pi }^2}
%% Jotas
+\frac{1}{2f_\pi^4}(s-2M_K^2)^2\bar{J}_{KK}(s)
\\
&~+\frac{1}{288f_\pi^4}
\left( 60 \left( u \left( 2 u + t \right)+ 4 s  M_K^2 - 8 M_K^4 \right)\bar{J}_{KK}(u)
+2(9 u -8 M_K^2-M_\pi^2-3M_\eta^2)^2 \frac{\bar{J}_{\pi \eta}(u)}{3} \right.\\
&~\left.
+\left.(9 u -2 M_\pi^2-6M_\eta^2)^2 \bar{J}_{\eta \eta }(u)
+  3 \left( u \left( 11 u + 4 t -8 M_K^2\right)  - 8 \left( u + 2
t -4M_K^2\right)  M_\pi^2 \right) \bar{J}_{\pi \pi }(u)+
\left(u\leftrightarrow t\right)
\rule[.5cm]{0cm}{.2mm}\right)\right]_{\rm NLO}\\
&~+\dots \ .
\end{aligned}
\end{equation}
%%%%%%%%%%%%%%%%%%%%%%%%%%%%
Here $s,t,u$ denote the Mandelstam variables and ellipses denote the higher chiral orders not taken into account.  The tadpole integrals arising from, e.g., the wave function re-normalization procedure are denoted by $\mu_i$ and read
%%%%%%%%%%%%%%%%%%%%%%%%%%%%
\begin{equation}
\begin{aligned}[b]
%\label{eqn:tabpoles}
\mu_i=\frac{M_i^2}{32\pi^2f_0^2}\log\frac{M_i^2}{\mu^2}\,,
\end{aligned}
\end{equation}
%%%%%%%%%%%%%%%%%%%%%%%%%%%%
where $\mu$ is the renormalization scale and $f_0$ is the meson decay constant in the chiral limit. The latter is determined dynamically, from employed sets of LECs and $\{M_\pi,M_K,f_\pi,f_K\}$ (see discussion below).  

The finite parts of the meson-meson loop integrals $\bar J{..}(s)$ are given by
%%%%%%%%%%%%%%%%%%%%%%%%%%%%
\begin{align}
%\label{eqn:tabpoles}
\bar J_{PQ}(s)=\frac{1}{32\pi^2}
    \Bigg(
        2
        &+\Big(\frac{\Delta(M_P, M_Q)}{s}-\frac{\Sigma(M_P, M_Q)}{\Delta(M_P, M_Q)}\Big)\log\frac{M_Q^2}{M_P^2}
        \\\nonumber
        &-\frac{\nu(s,M_P,M_Q)}{s}\Big(
        \log\frac{\Delta(M_P, M_Q)+s+\nu(s, M_P, M_Q)}{\Delta(M_P, M_Q)-s-\nu(s, M_P, M_Q)} -
        \log\frac{\Delta(M_P, M_Q)-s+\nu(s, M_P, M_Q)}{\Delta(M_P, M_Q)+s-\nu(s, M_P, M_Q)}
        \Big)
    \Bigg)\,,
\end{align}
%%%%%%%%%%%%%%%%%%%%%%%%%%%%
where $\nu(s,M_P,M_Q)= \sqrt{(s-(M_P + M_Q)^2)(s - (M_P - M_Q)^2)}$, $\Delta(M_P,M_Q)=M_P^2-M_Q^2$ and $\Sigma(M_P,M_Q)=M_P^2+M_Q^2$. The corresponding formula for equal masses simplifies to 
%%%%%%%%%%%%%%%%%%%%%%%%%%%%
\begin{equation}
\begin{aligned}[t]
%\label{eqn:tabpoles}
\bar J_{PP}(s)=\frac{1}{16\pi^2}
    \Bigg(
        2+\sigma(s)\log\frac{\sigma(s)-1}{\sigma(s)+1}
    \Bigg)
        ~~~\text{with}~~~
        \sigma(s)=\sqrt{1-4M_P^2/s}
    \,,
\end{aligned}
\end{equation}
%%%%%%%%%%%%%%%%%%%%%%%%%%%%
and resembles the self energy part $\rho(s)=16\pi\bar J_{KK}(s)$ used in the main part of the manuscript.

For the purpose of the present work, these formulas are projected to the $S$-wave. This implies a factor $N=2$ for the two identical kaons~\cite{GomezNicola:2001as}. The amplitude is subsequently back-transformed to the (two-body) plane-wave basis as required by the form of the quantization condition in Eq.~\eqref{eq:3bqc}. 

There are three mass relations and three relations for the decay constants~(see, e.g., Refs.~\cite{GomezNicola:2001as, Gasser:1984gg}) setting these physical quantities in relation to the tree level quantities $f_0,\,M_{0\pi},\,M_{0K}$, and $\, M_{0\eta}$ at NLO (we also replace $M_{0\eta}$ using the Gell-Mann--Okubo relation). For the chiral predictions, we follow two strategies, {\bf S1}: 
The tree level quantities are determined at the physical point from $M_\pi,\,M_K,\,M_\eta,\,f_\pi,f_K,\,f_\eta$. Then, $f_\pi$ is obtained by replacing tree level masses by the meson masses on the lattice in the relation
\begin{equation}
f_\pi=f_0\left[1-2\mu_\pi-\mu_K+\frac{4M_{0\pi}^2}{f_0^2}\left(L_4^r+L_5^r\right)+\frac{8M_{0K}^2}{f_0^2}\,L_4^r\right]\,.
\end{equation}
For strategy {\bf S2}: The tree level quantity $f_0$ is directly inferred from the available lattice information, in the present cases (GWUQCD and NPLQCD) $M_\pi,\,M_K,\,f_\pi,$ and $f_K$. We consider this method more reliable as one source of extrapolation uncertainty is removed. 

We replace also the $\eta$-mass in Eq.~(\ref{eq:amp4kch}) using the Gell-Mann--Okubo relation. All discussed replacements only lead to $\mathcal{O}(p^6)$ effects and are, thus, consistent to the chosen chiral order. Note also that the scattering amplitudes are regularization ($\mu$) independent, which implies that  the LECs are only defined at the given scale for which we choose the same value of $\mu=770$~MeV as in Ref.~\cite{Molina:2020qpw} to be able to use their LECs for predictions. Note that we cannot use the LECs of Ref.~\cite{Nebreda:2010wv} because they correspond to chiral amplitudes formulated in terms of $f_\pi,\,f_K,\,f_\eta$ as explained in Ref.~\cite{Pelaez:2004xp}. We emphasize that effects due to change of the renormalization
scale have been studied thoroughly in the two-flavour case in Ref.~\cite{Mai:2019pqr}, where they have been found subdominant to the other effects, e.g., due to pseudo-scalar masses, decay constants and LECs.

While the sets of available SU(2) LECs produce very consistent predictions of the three-pion spectrum~\cite{Culver:2019vvu, Mai:2019fba}, the eight SU(3) LECs are less well determined; we chose the LECs of Ref.~\cite{Molina:2020qpw} (fit 4) for this study because the pertinent fit includes lattice data; if one uses the older LECs of Ref.~\cite{GomezNicola:2001as} obtained from only fits to experimental data, the scattering length turns out to be about 25\% smaller than the one shown in Fig.~\ref{fig:scatlength}, in contradiction with the NPLQCD results~\cite{Beane:2006kx}; likewise, the three-body ground state energy shift gets about 25\% smaller, in contradiction with the lattice data.

%%%%%%%%%%%%%%%%%%%%%%%%%%%%%%%%%%%%%%%%%%%%%%%%%%%%%%%%%%%%%
\section{Lattice energies}\label{app:LQCD}
%%%%%%%%%%%%%%%%%%%%%%%%%%%%%%%%%%%%%%%%%%%%%%%%%%%%%%%%%%%%%
Here we tabulate the lattice energy levels extracted from the ensembles in Table~\ref{table:gwu_lattice} in the main text.  For the $315\MeV$ ensemble ($\mathcal{E}_1$), triple exponential fits were performed as in Ref.~\cite{Culver:2019vvu}.  Due to an increase in the noise on the $220\MeV$ ensemble ($\mathcal{E}_4$), only single exponential fits were performed.  The energies are tabulated in Table~\ref{table:gwu_energies}. The final column shows the energies predicted from SU(3) IAM using lattice decay constants and masses from Table~\ref{table:gwu_lattice} of the main text.

\begin{table}[h]
    \centering
\begin{ruledtabular}
\begin{tabular}{@{}*{13}{>{$}l<{$}}@{}}    
        {\rm Ensemble} & {\rm Irrep} & E_\text{lat}/M_K & E_\text{pred}/M_K\\
        \midrule
        \mathcal{E}_1 & A_{1u} & 3.037(4) & 3.0333\\
                      &        & 3.656(5) & 3.6706\\ 
                      & E_{u}  & 3.610(5) & 3.6134\\
        \midrule
        \mathcal{E}_4 & A_{1u} & 3.035(4) & 3.0367\\
                      &        & 3.697(4) & 3.7198\\
                      & E_{u}  & 3.650(5) & 3.6578
    \end{tabular}
    \end{ruledtabular}
    \caption{Energy levels in different irreps of the cubic group, as described in the main text.}
    \label{table:gwu_energies}
\end{table}

\end{onecolumngrid}

\end{document}